# Life 2.0: A Scalable Distributed Space-Telescope Array for Biosignature Spectroscopy

Jian Ge, Zhangqi Dang, Ziru Zhang, Chenxu Gao, Shijie Ke, and Ziyang Zhang, Shanghai Astronomical Observatory, CAS; Rong Shu and Wen Chen, Innovation Academy for Microsatellites, CAS; Jie Yin, Yunzhou Zhu, Leiming Lei, and Zhongming Chen, Shanghai Institute of Ceramics, CAS; Jiancheng Ji and Xiangsen Tian, Shanghai YUM Oe Tech Co., Ltd.; Jun Yang and Xinyi Song, Peking University; Rafael Luque, Instituto de Astrofisica de Andalucia; Enric Palle, Instituto de Astrofísica de Canarias, Spain

Corresponding author: astrojge@gmail.com

## ABSTRACT

Answering the question "Are we alone?" ultimately requires atmospheric spectroscopy of nearby terrestrial planets. For an Earth–Sun analog, even the strongest transmission signals are expected to be only of order 1 part per million (ppm). Unlike short-period planets, however, Earth 2.0 planets transit only about once per year, making single-transit sensitivity, rather than the accumulation of many repeated observations, the fundamental design driver. Life 2.0 is a scalable space-mission concept that bridges the discovery of Earth 2.0 candidates by transit surveys such as PLATO and the Earth 2.0 (ET) mission with atmospheric characterization and biosignature assessment. The baseline architecture comprises 900 one-meter space telescopes, each equipped with a high-throughput Waveguide Integrated Miniature Spectrograph (WIMS) and an ultra-low-read-noise CMOS detector. After independent calibration, spectra acquired simultaneously during a transit are statistically combined, providing the photon-collecting capability of an approximately 30-m aperture at the selected spectral resolution while retaining the advantages of a modular architecture. The baseline 0.2–1.05 μm wavelength range covers key atmospheric diagnostics including $O_3$, $O_2$, $H_2O$, Rayleigh scattering, and other molecular species, with potential extension into the infrared as detector technologies mature. Prototype Waveguide Spectral Lens (WSL) devices have demonstrated 40–66% throughput at resolving powers from $R \approx 200$ to $R \approx 20,000$, while lightweight silicon-carbide mirrors and sub-electron-noise CMOS detectors provide a practical path toward large-scale replicated production. Like all transmission spectroscopy missions, Life 2.0 must address detector systematics, instrument stability, and stellar variability. Rather than assuming these limitations disappear, the mission builds upon the rapidly advancing calibration, detector characterization, and data-analysis techniques pioneered during the JWST era. Life 2.0 offers a scalable alternative to a monolithic 30-m-class space telescope and a staged technological pathway toward atmospheric spectroscopy and biosignature searches for nearby Earth-like planets.

## 1. INTRODUCTION

"Are we alone in the Universe?" is one of humanity's most fundamental scientific questions. The observational path toward an answer begins with the discovery of terrestrial planets in the habitable zones of nearby stars and culminates in measurements of atmospheric composition, climate, and potential biosignatures [1–3]. More than 6,000 exoplanets are now known, and space-based transit surveys have transformed the field from isolated discoveries into population science. Nevertheless, a bona fide Earth 2.0—defined here as a 0.8–1.25 Earth-radius planet in the habitable zone of a Sun-like star with an approximately one-year orbital period—has not yet been established as a well-characterized population [4].

China's Earth 2.0 (ET) space mission is designed to discover Earth 2.0 planets and measure their occurrence rate through long-duration, continuous, ultra-high-precision photometry [5–9]. ET, PLATO, and complementary space-based photometric surveys, particularly those targeting nearby bright stars, are expected to provide the high-value target lists required for the next stage of exoplanet science. A complete scientific capability chain must therefore extend from planet discovery to atmospheric characterization and, ultimately, to robust biosignature assessment. Life 2.0 is conceived as the natural successor to this discovery phase, providing a modular platform for ultra-stable, multi-band transmission spectroscopy of nearby transiting Earth 2.0 candidates.

The rarity of observable transits fundamentally distinguishes the atmospheric characterization of Earth 2.0s from that of most currently known exoplanets. For short-period planets, transmission spectra can be improved by co-adding observations from dozens of transit events. For true Earth 2.0s, however, the orbital period is approximately one year, allowing only a single observable transit annually, with additional losses possible because of mission scheduling, spacecraft operations, or other practical constraints. Consequently, the primary design driver becomes the precision achieved during an individual transit rather than the cumulative precision obtained from repeated observations. This requirement strongly favors an observatory capable of delivering an exceptionally large instantaneous photon-collecting capability, motivating the distributed-aperture architecture adopted by Life 2.0.

JWST has demonstrated the transformative scientific power of high-precision transmission spectroscopy through detections of $H_2O$, $CO_2$, $SO_2$, $CH_4$, and other molecular species in exoplanet atmospheres [11–14]. For an Earth–Sun analog, however, one atmospheric scale height produces a transmission signal of only about 0.2 ppm, while even the strongest molecular absorption bands are expected to reach only ~1 ppm. These amplitudes are well below the tens-of-ppm precision typically achieved in individual spectral channels for current observations. Achieving such sensitivity requires not only a very large effective collecting area, but also an instrument architecture designed from the outset to minimize correlated systematic errors at the ppm and sub-ppm level.

Importantly, Life 2.0 does not eliminate the fundamental challenges already encountered in JWST transmission spectroscopy. Detector systematics, pointing jitter, thermal drifts, wavelength calibration stability, stellar activity, granulation, starspots, faculae, and other sources of astrophysical and instrumental variability will remain important limitations for observations of terrestrial exoplanets. Fortunately, the rapid maturation of transmission spectroscopy during the JWST era is producing powerful new calibration strategies, detector characterization methods, systematic-noise models, stellar activity mitigation techniques, and advanced Bayesian and machine-learning retrieval frameworks. Life 2.0 is designed to leverage these developments while incorporating lessons learned from JWST into both its hardware architecture and end-to-end data analysis pipeline. Rather than beginning from scratch, the mission builds upon the rapidly advancing foundation established by the current generation of space-based exoplanet spectroscopy.

Conventional approaches to achieving the required sensitivity generally rely on a single extremely large space telescope, high-contrast direct imaging, or a complex space interferometer [15,16]. Life 2.0 instead explores a complementary distributed architecture in which hundreds of identical small telescopes observe the same transit simultaneously, while their independently calibrated spectra are statistically combined. This approach exchanges the complexity and risk of a monolithic observatory for modular manufacturing, parallel integration and testing, graceful degradation, and a scalable development path from technology demonstrations to a full-scale observatory capable of single-transit atmospheric characterization of nearby Earth 2.0 planets.

## 2. SCIENTIFIC MOTIVATION AND MEASUREMENT REQUIREMENTS

### 2.1 From Earth 2.0 discovery to biosignature testing

Ultra-high-precision space photometry, pioneered by Kepler and extended by TESS, CHEOPS, PLATO, and ET, provides the most direct route to statistically useful samples of small transiting planets [4-10]. Discovery missions determine orbital periods, radii, ephemerides, and prioritized target lists. A follow-up observatory must then obtain repeated spectra over many transits, combine measurements across time and across telescope modules, and jointly model stellar variability, transit geometry, instrumental drifts, and planetary atmospheric signals. This discovery-to-characterization chain creates a closed scientific pathway from candidate identification to atmospheric retrieval and biosignature assessment.

### 2.2 Amplitude of an Earth 2.0 atmospheric signal

For transmission spectroscopy, the approximate contribution of one atmospheric scale height to the transit depth is

$$\delta_h \approx 2R_pH/R_\star^2,$$

where $R_p$ is the planetary radius, H is the atmospheric scale height, and $R_\star$ is the stellar radius. For the Earth-Sun system ($R_p = R_\oplus$ and H ≈ 8 km), one scale height contributes approximately 0.2 ppm. Strong molecular bands may span several effective scale heights and approach the 1 ppm level, although clouds, atmospheric composition, stellar activity, and the limited number of observable transits can reduce the recoverable signal. A long-term differential spectrophotometric stability goal of approximately 1 ppm is therefore appropriate at the system level, while the error budget should preserve sensitivity to sub-ppm spectral structure.

### 2.3 Photon statistics and the systematic-error floor

Under photon-limited conditions, the statistical uncertainty scales approximately as

$$\sigma_{photon} \approx 1/\sqrt{N_{photon}} \propto 1/(D\sqrt{t}),$$

where $N_{photon}$ is the detected photon count, $D$ is the aperture diameter of one telescope, and $t$ is the accumulated integration time. For $N$ identical telescopes, the equivalent photon-collecting diameter is $D_{eff} = \sqrt{N}$ D. Thus, 900 one-meter telescopes provide the same geometric collecting area as a 30-m aperture. This equivalence refers to photon count and photon-limited signal-to-noise ratio at the same throughput, spectral resolution, and integration time; it does not imply the angular resolution of a filled 30-m aperture. The distributed array therefore targets 30-m-class spectroscopic sensitivity while retaining a modular and scalable architecture.

The total uncertainty, however, is determined by both photon noise and residual systematics:

$$\sigma_{total}^2 \approx \sigma_{photon}^2 + \sigma_{sys}^2.$$

As the photon-noise term is reduced through collecting area and repeated transits, residual systematics increasingly determine the achievable precision. Uncorrelated module-level errors may average down, but common-mode and time-correlated errors do not generally follow the $\sqrt{N}$ scaling. Increasing collecting area alone therefore provides little benefit unless detector nonlinearity, electronic crosstalk, gain variations, thermo-mechanical drift, stray light, wavelength-calibration errors, pointing-dependent coupling, stellar variability, and residual pipeline systematics are controlled through an end-to-end error budget over multi-year operations.

### 2.4 Key technology requirements and validation pathway

1) Ultra-stable spectrographs with continuous monitoring of wavelength, throughput, and line-spread-function drift;
2) Precision calibration of detector systematics, including nonlinearity, crosstalk, persistence, interpixel response, and gain drift;
3) Thermo-mechanical stability, stray-light suppression, stable fiber coupling, and repeatable optical-path control;

4) Reference sources and wavelength-calibration systems stable enough to support ppm-level differential measurements;
5) Array-level synchronization, co-pointing, module-to-module calibration, and robust spectral registration and co-addition; and
6) End-to-end simulators, validation experiments, and systematic-error models explicitly tied to the 1 ppm stability requirement.

The near-term Ultra-high Precision CMOS Photometer (UPCP) provides an integrated validation platform for these capabilities. A vacuum-compatible UPCP testbed links detector calibration, optical and thermal stability, reference-source metrology, controlled pointing disturbances, and end-to-end data analysis. In the near term, it can validate observing and calibration strategies for ET-era targets; in the longer term, it serves as an engineering pathfinder for a distributed Life 2.0 observatory.

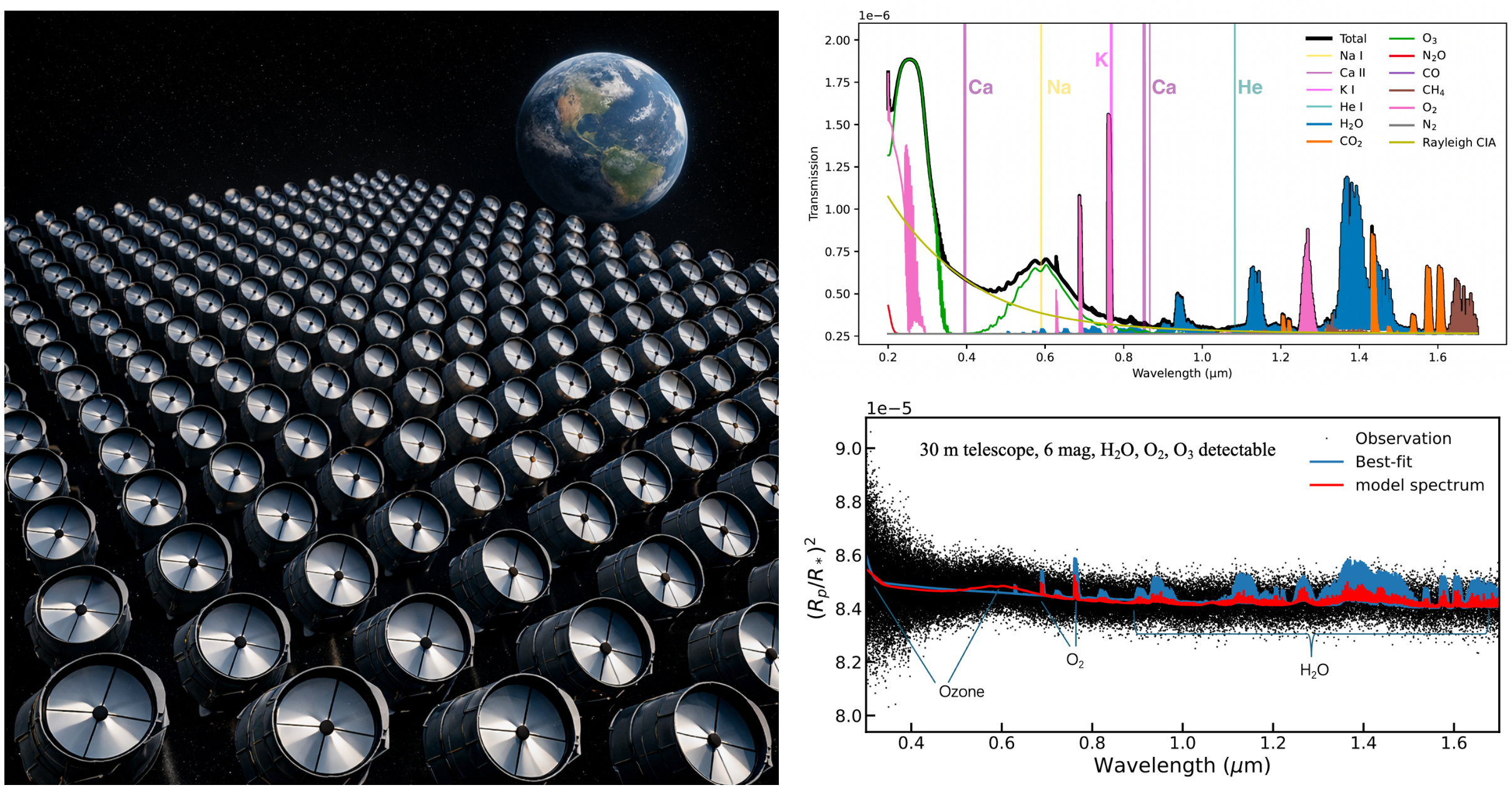

*Figure 1. Life 2.0 mission concept and simulated biosignature spectroscopy. The distributed array observes a common target with many one-meter telescope modules. The example spectra illustrate UV-optical atmospheric signatures and the simulated recovery of $O_3$, $O_2$, and $H_2O$ after calibrated spectral co-addition.*

## 3. MISSION ARCHITECTURE

The baseline Life 2.0 architecture employs 900 one-meter space telescopes that observe a common target star simultaneously (Figure 1). The concept extends the distributed-aperture philosophy of the Large Fiber Array Spectroscopic Telescope (LFAST) [17] from ground-based high-resolution spectroscopy to a space platform optimized for ultra-stable transit spectrophotometry. Each telescope module feeds a compact WIMS and an ultra-low-read-noise detector. For $N$ identical apertures of diameter $D$, the equivalent photon-collecting diameter $D_{eff} = \sqrt{N}\,D$; with $N = 900$ and $D = 1$ m, the total geometric area equals that of a 30-m aperture. The architecture is designed to deliver the photon statistics required for weak atmospheric signatures while keeping each module independently manufacturable, testable, calibratable, and suitable for parallel production. Modules can be monitored and excluded individually without interrupting the full array, providing graceful degradation.

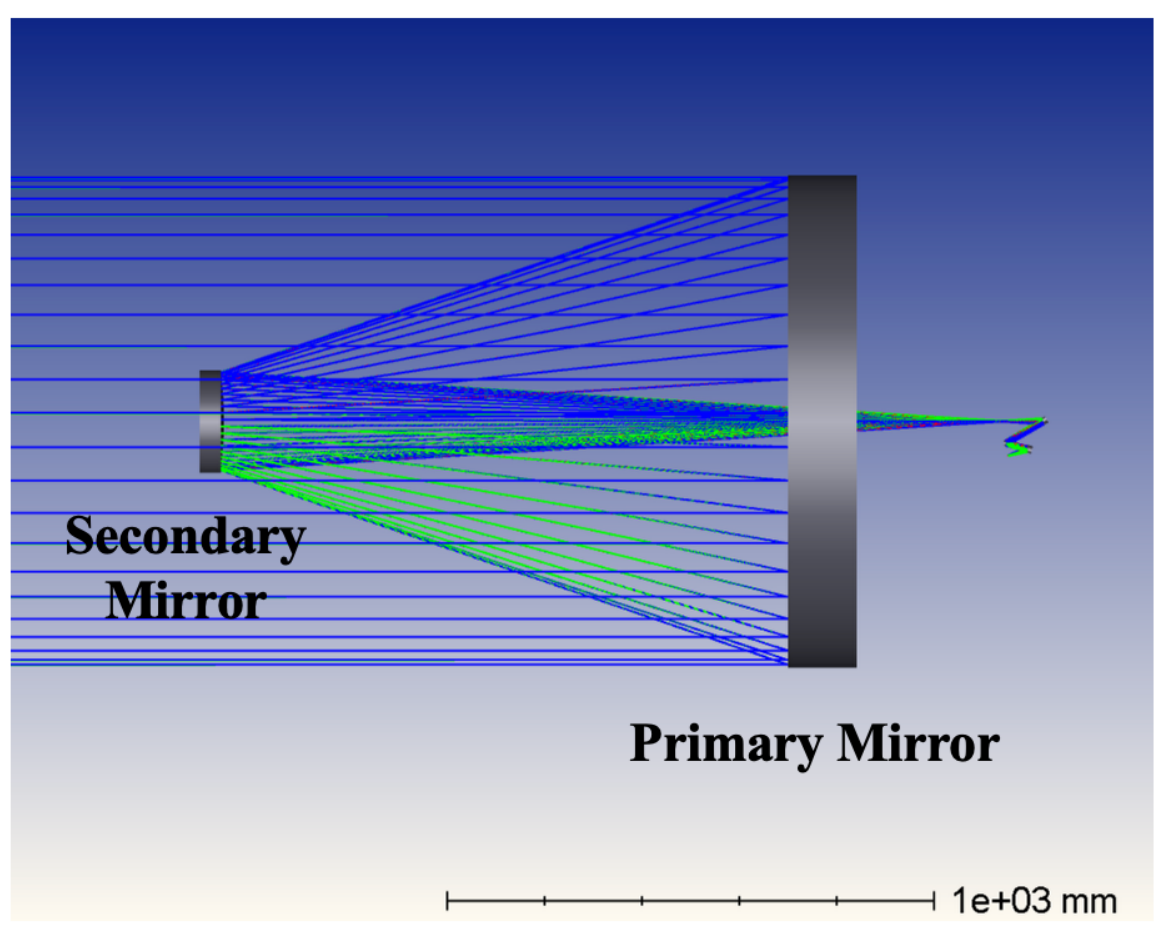

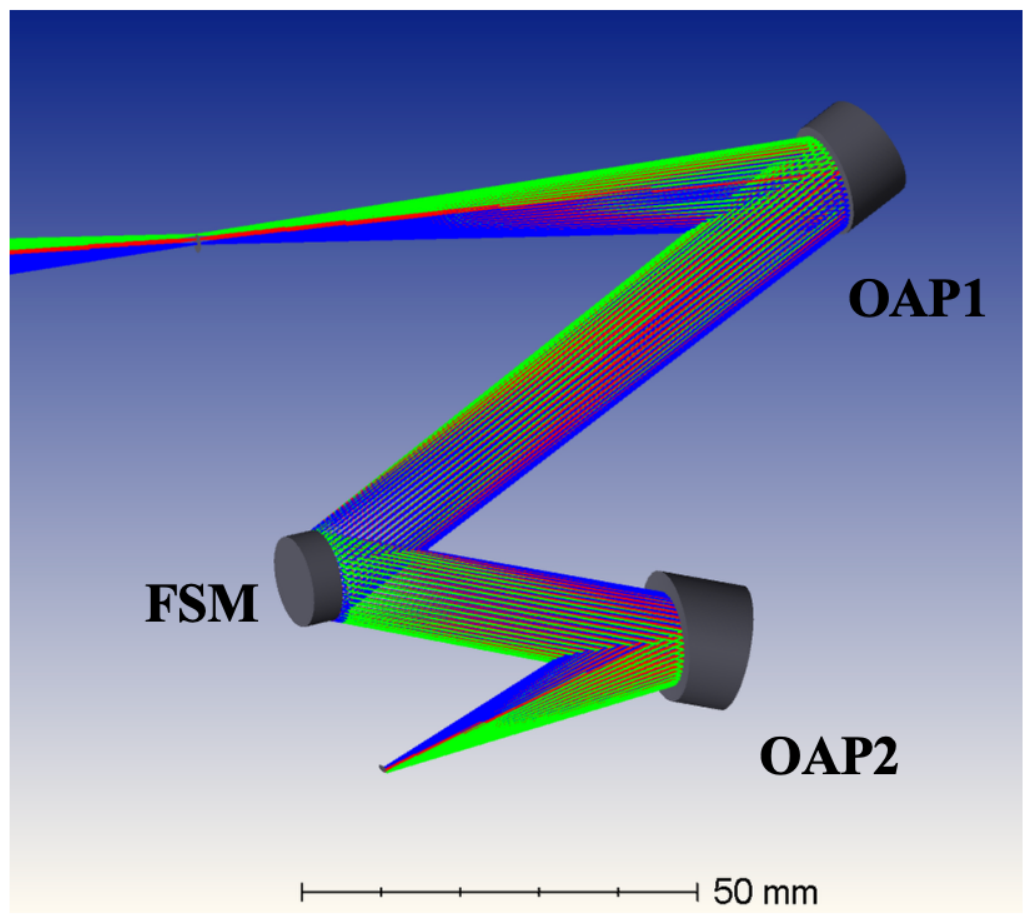

*Figure 2. Optical design of one modular 1-m Life 2.0 telescope. An array of 900 such modules would collectively provide the geometric collecting area of a 30-m aperture.*

Figure 2 illustrates the optical design of a modular 1-m telescope for the Life 2.0 array. Because each module is dedicated to continuous observations of a bright nearby target, the optical system can use a compact *f*/8 Cassegrain configuration with a parabolic primary mirror, minimizing complexity, mass, and cost while maintaining diffraction-limited image quality. At the Cassegrain focus, an off-axis parabola re-images the pupil onto a folding mirror that serves as the fast-steering element. A second off-axis parabola converts the *f*/8 beam to *f*/4 for efficient coupling into a single-mode fiber. The fiber feeds a high-throughput WIMS, which records the time-dependent stellar spectrum before, during, and after planetary transits. The pupil-plane folding mirror is mounted on a high-bandwidth tip-tilt actuator designed to stabilize the line of sight to approximately 0.01 arcsec, about one-tenth of the diffraction-limited image scale at visible wavelengths. Stable beam injection and fiber coupling are essential because coupling fluctuations can otherwise mimic or obscure ppm-level transit signals.

The baseline mission-level science range extends from approximately 0.2 to 1.05 µm, providing access to $O_3$, $O_2$, $H_2O$, Rayleigh scattering, alkali lines, aerosols, and other atmospheric diagnostics (Figure 1). In practice, this broad range would be partitioned among UV, visible, and near-infrared channels optimized for detector response, fiber transmission, and photonic-material performance; the fused-silica WSL prototypes described below validate selected visible and near-infrared wavelengths. The architecture can be extended to longer wavelengths as sufficiently low-noise infrared detectors mature. A modular array is particularly well suited to repeated monitoring because spectra can be combined across telescopes and transits while retaining module-level data for systematic-error diagnosis.

## 4. DESIGN PARAMETERS AND SYSTEM APPROACH

The Life 2.0 concept is built around four coupled elements: a lightweight 1-m telescope module, a high-throughput WIMS, an ultra-low-read-noise CMOS detector, and an end-to-end calibration architecture capable of preserving ppm-level differential stability. Table 1 summarizes representative parameters for the current mission concept; these values are design goals or demonstrated component values rather than a completed flight-system specification.

### 4.1 Lightweight one-meter telescope

Each optical module uses a compact 1-m telescope with folded, diffraction-limited optics. Silicon carbide (SiC) is attractive because of its high stiffness-to-mass ratio, thermal stability, and compatibility with lightweight mirror structures. Additive manufacturing followed by precision finishing and metrology offers a route to reducing recurring mass, fabrication time, and cost, all of which become decisive when hundreds of nominally identical apertures must be produced and qualified.

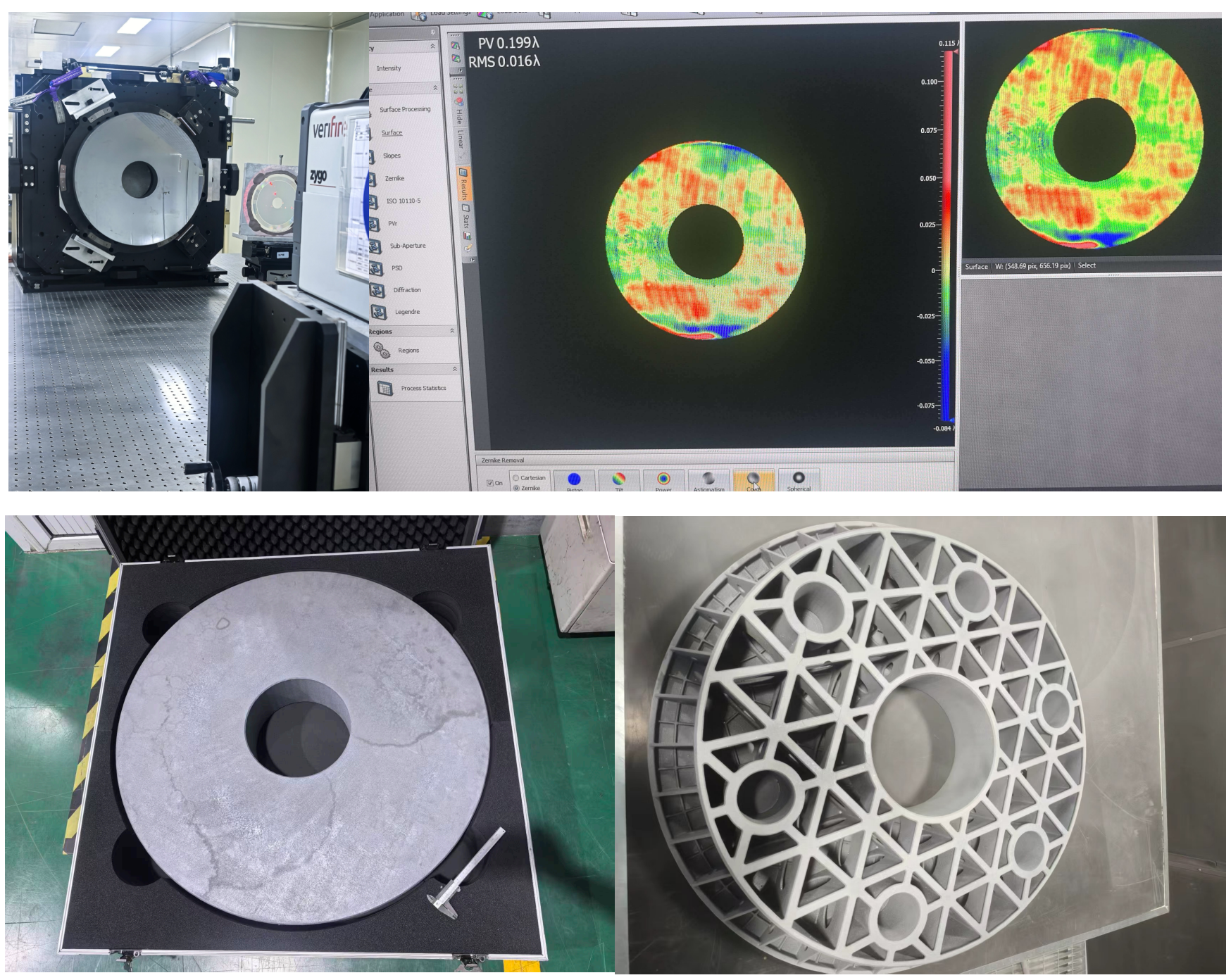

*Figure 3. Lightweight silicon-carbide (SiC) mirror technology for Life 2.0 telescope modules. The upper panels show fabrication, polishing, and surface metrology of a lightweight 0.5-m SiC mirror. The lower panels show an additively manufactured 1-m SiC mirror and its integrated lightweight back-support structure, illustrating a scalable path toward low-mass optics for a distributed space-telescope array.*

Figure 3 summarizes the current SiC mirror development, including fabrication, polishing, surface metrology of a 0.5-m lightweight mirror, and an additively manufactured 1-m lightweight mirror structure. Together with the off-axis relay optics and fast steering concept in Figure 2, these demonstrations provide an initial manufacturing basis for a replicated telescope module. Flight development will additionally require environmental qualification, long-term dimensional-stability measurements, coating validation, and a production-level acceptance-test strategy.

## 5. WAVEGUIDE-INTEGRATED MINIATURE SPECTROGRAPH

A central enabling technology for Life 2.0 is the WIMS. The array requires a spectrograph that is compact, stable, low in mass, and suitable for replicated production. Rather than assigning a conventional bulk-optics spectrograph to every telescope, Life 2.0 adopts the WIMS, a next-generation implementation of the WSL concept. The wafer design and some of the fabricated WSL chips are shown in Figure 4. By integrating the spectroscopic functions onto a photonic chip and eliminating the external relay optics, bulk disperser, and imaging optics used in the current WSL prototype, WIMS dramatically reduces instrument size, mass, alignment complexity, and mechanical sensitivity while preserving high optical throughput. This highly integrated architecture is particularly well suited for the mass production of hundreds of identical spectrographs required by the Life 2.0 distributed telescope array.

Prototype fused-silica WSL devices have been fabricated and tested at resolving powers near R ≈ 150-200, R ≈ 2,000, and R ≈ 20,000, with measured device throughputs of 40-66% in the configurations summarized in Table 2 [18,19]. Multiple resolutions support complementary observing modes: low resolution maximizes sensitivity to broad molecular bands, medium resolution supports atmospheric retrieval and wavelength calibration, and high resolution enables targeted line diagnostics and cross-correlation techniques. A flight WIMS would use the resolution and channelization that maximize biosignature information under the mission-level photon and stability budgets.

| Subsystem / parameter | Representative value | Purpose |
|---|---|---|
| Telescope array | 900 × 1-m apertures | Replicated 30-m-class collecting area |
| Equivalent aperture | ~30-m equivalent collecting diameter | Equal photon count at fixed throughput and R |
| Science band | 0.2-1.05 μm in optimized channels | Atmospheric and biosignature diagnostics |
| Spectrograph | WIMS photonic spectrograph | Compact, high-throughput, replicable |
| Resolving power, R | R ≈ 150-200, 2,000, and 20,000 | Broad bands, retrievals, and line diagnostics |
| Detector | Back-illuminated CMOS; ~0.4 $e^-$ median | Low read-noise penalty at high cadence |
| Primary mirror | Lightweight additively manufactured SiC | Scalable low-mass optics |
| Systematic precision floor | ~1 ppm long-term stability goal | Earth-analog atmospheric signals |
| Calibration architecture | Detector, module, and array levels | Controls module- and array-level systematics |

*Table 1. Representative design parameters and technology targets for the Life 2.0 distributed-array concept.*

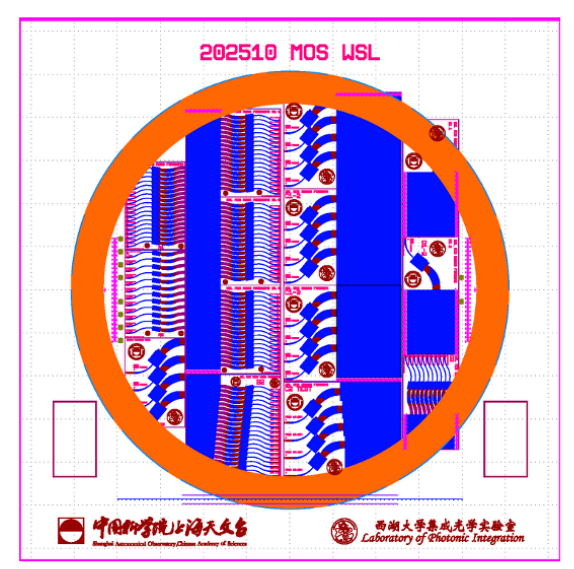

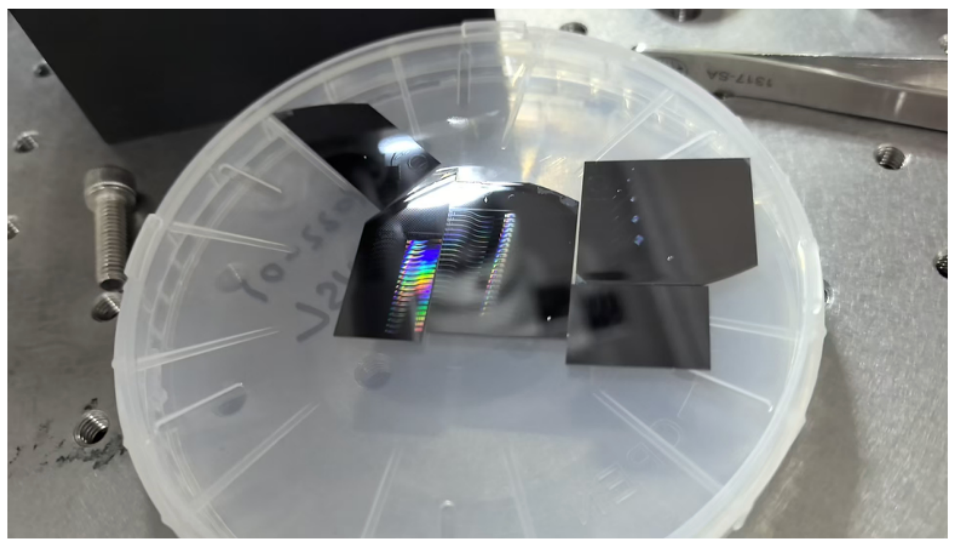
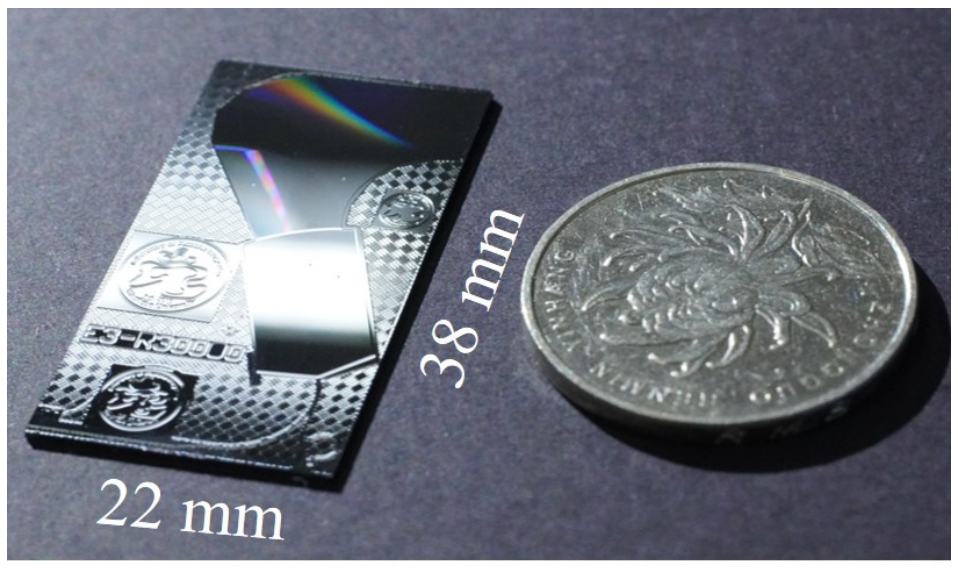

*Figure 4. Waveguide-integrated spectrograph technology. From left to right, the panels show a WSL chip design, fabricated low- and medium-resolution devices, and a fabricated high-resolution WSL device, illustrating the scalability of the waveguide concept across resolving power.*

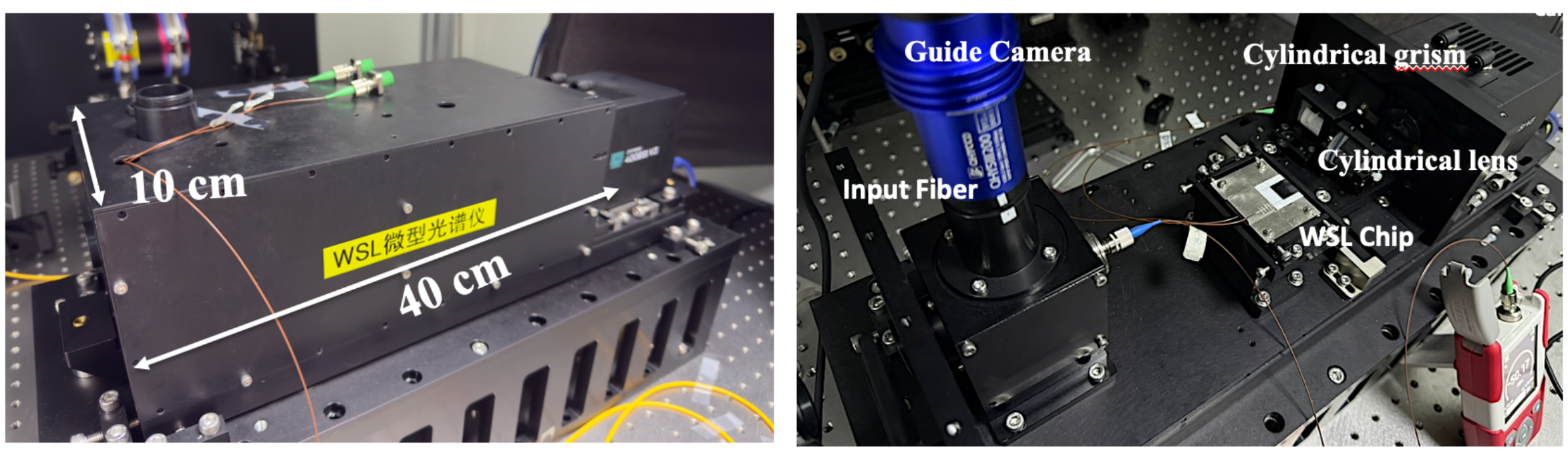

*Figure 5. Integrated WSL prototype spectrograph. The left panel shows the assembled instrument, and the right panel shows its internal optical layout, including the input fiber, WSL photonic chip, cylindrical grism, cylindrical lens, guide camera, and detector camera. The compact optical train demonstrates the high level of integration required for replicated WIMS production.*

To advance the photonic spectrograph technology required by Life 2.0, an integrated WSL prototype spectrograph was developed during 2024-2026. Figure 5 shows the assembled instrument and its internal optical configuration. The prototype has undergone extensive laboratory characterization and on-sky testing on the Lijiang 1.8-m telescope equipped with adaptive optics. Integrated WSL spectrographs at R ≈ 200 and R ≈ 2,000 have been demonstrated, establishing the practical feasibility of compact photonic spectrographs for astronomical observations. These results form an important technology foundation for the replicated WIMS units proposed for Life 2.0. Detailed device characterization and on-sky performance are presented in companion SPIE papers [18,19].

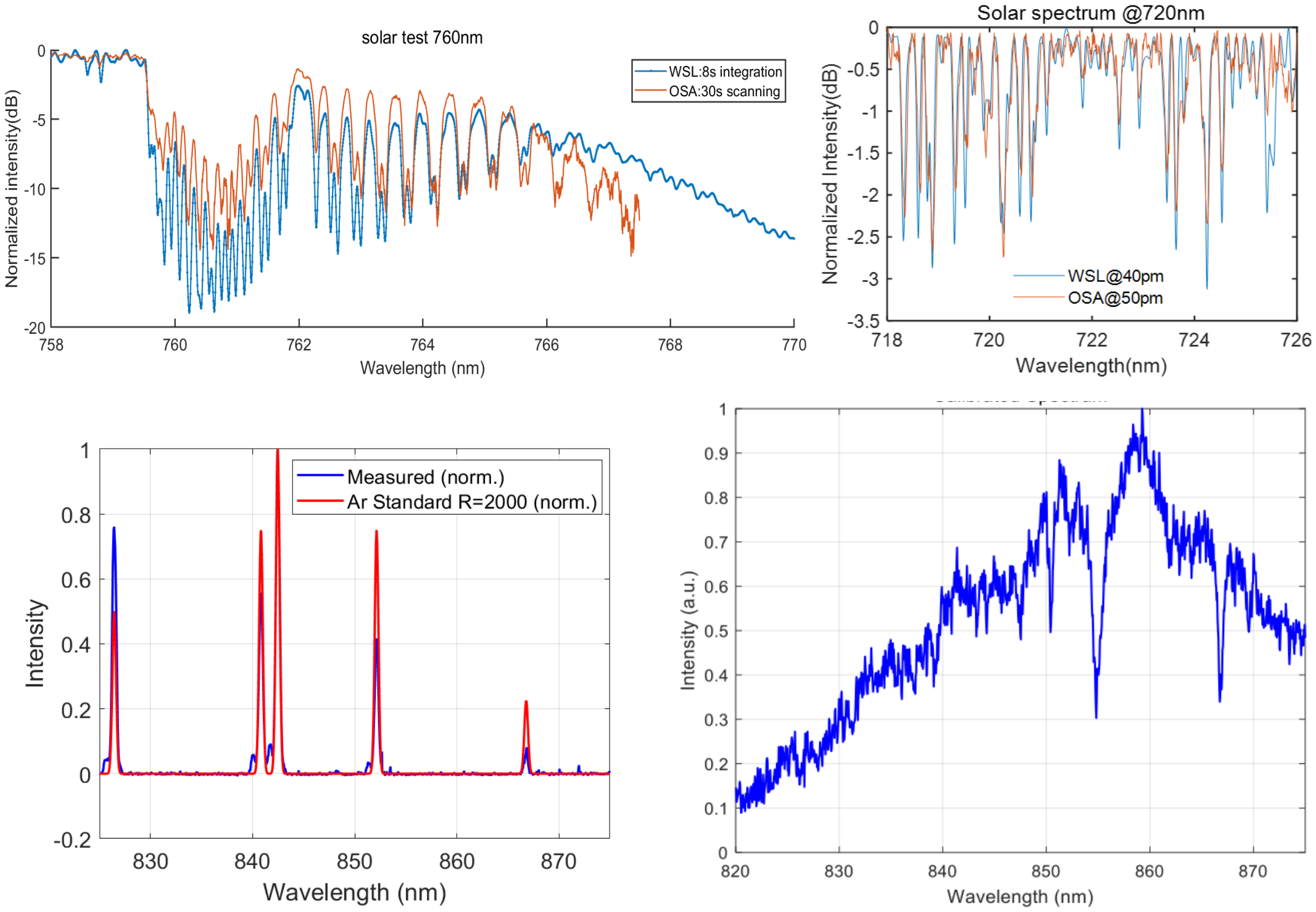

*Figure 6. Laboratory and on-sky demonstrations of WSL spectrograph performance. The top panels show R ≈ 20,000 measurements of telluric $O_2$ and $H_2O$ absorption imprinted on solar spectra. The lower-left panel shows a Neon wavelength-calibration spectrum obtained with the R ≈ 2,000 integrated WSL prototype, and the lower-right panel shows an R ≈ 2,000 spectrum of α Orionis (Betelgeuse) acquired with the same instrument on the adaptive-optics-equipped Lijiang 1.8-m telescope.*

Laboratory and on-sky demonstrations of WSL technology provide important risk reduction for Life 2.0. Figure 6 presents high-resolution (R ≈ 20,000) measurements of telluric $O_2$ and $H_2O$ absorption imprinted on solar spectra, together with an R ≈ 2,000 Ne wavelength-calibration spectrum and a first-light stellar spectrum obtained with the integrated prototype [18,19]. These experiments demonstrate that WSL devices can resolve narrow atmospheric and stellar features over a wide range of resolving powers and can operate as practical astronomical spectrographs. The results establish component and instrument feasibility, while the next development step is to demonstrate long-term spectrophotometric stability, repeatability, and environmental robustness under flight-relevant conditions.

### 5.1 WSL throughput demonstrations

Table 2 summarizes measured throughputs of 40-66% for representative fused-silica WSL devices at visible and near-infrared wavelengths. These device-level results are encouraging because they show that waveguide dispersion need not impose a severe throughput penalty. For context, the GTC reports wavelength- and grism-dependent end-to-end photon-detection efficiencies for the conventional OSIRIS spectrograph (https://www.gtc.iac.es/instruments/osiris/). A direct numerical comparison should be made cautiously, however, because the WSL measurements and telescope-level OSIRIS measurements do not necessarily include the same coupling, optical, detector, and aperture losses. The mission-relevant next step is therefore an end-to-end WIMS throughput measurement that includes fiber injection, detector quantum efficiency, and stability over time. Nevertheless, the demonstrated combination of high device throughput, compactness, and scalability strongly supports WSL/WIMS technology for photon-limited exoplanet spectroscopy in a distributed array.

| Resolving power, R | Central wavelength (nm) | Throughput |
|---|---|---|
| 19,000 | 760 | 40% |
| 2,057 | 800 | 48% |
| 1,955 | 800 | 54% |
| 1,600 | 1550 | 66% |
| 150 | 800 | 49% |
| 177 | 1300 | 50% |

*Table 2. Measured device throughput for representative WSL configurations. Values were obtained at different wavelengths and resolving powers and should not be interpreted as end-to-end telescope efficiencies.*

## 6. ULTRA-LOW-READ-NOISE DETECTORS AND INTEGRATED SYSTEM TESTING

Ultra-low detector noise is a key enabling technology for the Life 2.0 distributed architecture. The mission seeks 30-m-class photon-collecting capability by combining spectra acquired simultaneously with 900 independent 1-m telescopes. To realize the expected $\sqrt{N}$ gain, the uncorrelated detector and instrument noise in each module must be subdominant to its photon noise, while correlated errors shared among modules must be controlled below the final differential-stability requirement. Under these conditions, calibrated co-addition of the 900 spectra can achieve the photon-limited signal-to-noise ratio of a 30-m aperture at the same throughput, resolving power, and integration time. The spectra are combined statistically rather than through coherent optical beam combination.

Life 2.0 therefore targets the 0.3–0.4 $e^-$ read-noise regime using state-of-the-art back-illuminated scientific CMOS detectors. For example, the commercially available GSENSE6504BSI achieves a median read noise of approximately 0.4 $e^-$ in its ultra-low-noise mode (Figure 7). Such performance enables high-cadence, short-exposure observations of the mission's primary targets—nearby bright solar-type stars (typically $V \approx 6$ mag)—while introducing a negligible read-noise penalty and preserving photon-noise-limited performance. However, achieving the mission's ~1 ppm spectrophotometric precision requires more than ultra-low read noise. Other detector systematics, including dark current, persistence, nonlinearity, interpixel response variations, fixed-pattern noise, gain drift, and radiation-induced degradation, must also be calibrated and controlled within the overall system error budget. Combined with the high throughput of the WIMS photonic spectrographs and ultra-stable end-to-end calibration, these detectors provide a credible technological foundation for a distributed observatory capable of delivering the photon-limited spectroscopic sensitivity of a virtual 30-m-class space telescope.

To validate the complete instrument chain, a vacuum-compatible UPCP laboratory testbed is being developed for integrated stability testing rather than isolated component demonstrations. The platform can inject calibrated simulated stellar spectra and transit-like signals while controlling detector and telescope temperatures, illumination history, spectral format, pointing motion, fiber-coupling changes, optical-path drift, and other environmental variables. Repeated experiments will measure how component-level errors propagate into extracted time-series spectra, determine which errors are common-mode or wavelength dependent, and test whether calibration models remain valid over transit, campaign, and mission-representative timescales. The central validation metric is not a single short-term noise value, but repeatable recovery of known sub-ppm spectral signals after the complete calibration and extraction pipeline.

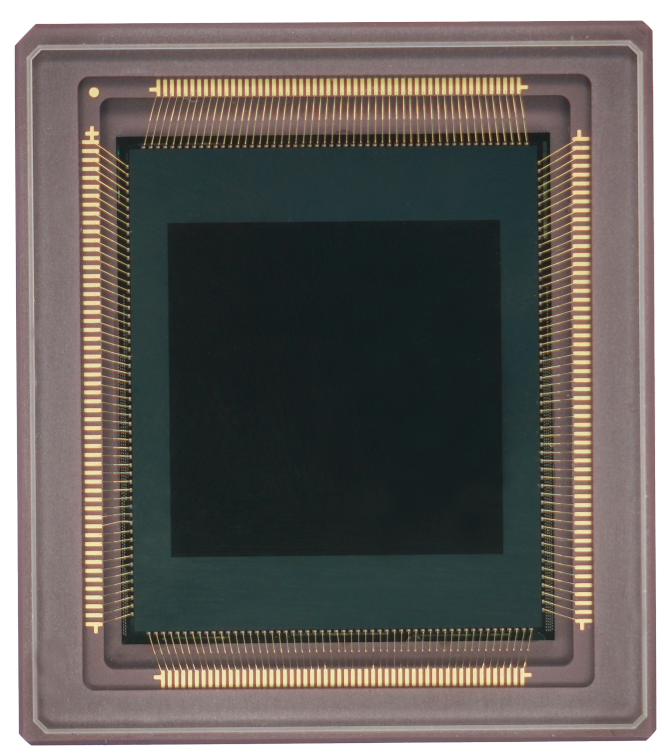

| Sensor | GSENSE6504BSI | GSENSE2020BSI |
| --- | --- | --- |
| Median Noise | 0.4 $e^-$ | 1.2 $e^-$ |
| Image | | |

*Figure 7. GSENSE6504BSI back-illuminated CMOS detector considered for the Life 2.0 WIMS. The left panel shows the detector package, and the right panel compares its approximately 0.4 $e^-$ median read noise with that of a representative scientific CMOS detector. This sub-electron performance is promising for high-cadence, photon-limited transit spectroscopy.*

At the observatory level, calibration must distinguish the planetary transit signal and stellar variability from both module-specific errors and instrumental disturbances correlated across the array. Stable reference sources, synchronized exposures, traceable wavelength grids, module-level quality metrics, and hierarchical spectral combination will be required. Replication is also an advantage: hundreds of nominally identical modules provide internal redundancy for identifying outliers, monitoring health, and separating reproducible astrophysical signals from hardware-specific behavior. This advantage is realized only if the calibration state, thermal history, pointing performance, and detector behavior of every module are continuously recorded. Array-level demonstrations with multiple synchronized modules are therefore an essential intermediate milestone before scaling to the full observatory.

## 7. SCIENCE CAPABILITY AND EXPECTED DEVELOPMENT PATH

The primary science objective is to detect and jointly interpret atmospheric features in nearby transiting Earth 2.0 candidates. UV-to-near-infrared measurements of $O_3$, $O_2$, $H_2O$, Rayleigh scattering, aerosols, and atomic species can constrain atmospheric composition, pressure structure, clouds, photochemistry, and surface-atmosphere context. No single species constitutes proof of life; the scientific goal is to evaluate combinations of gases in their planetary and stellar environments. Multi-resolution spectroscopy is valuable because broad molecular bands, narrow line complexes, cross-correlation diagnostics, and calibration references favor different resolving powers.

Scalability is the distinguishing feature of Life 2.0. The array avoids fabrication and deployment of a single 30-m-class monolithic telescope while preserving total photon-collecting area through calibrated spectral co-addition. It also offers operational redundancy: degraded modules can be identified and excluded without terminating the observatory, and a servicing-enabled architecture could permit later replacement or upgrade. The central engineering challenge is to make many independent modules behave as one precision spectrophotometric system through detector uniformity, synchronized timing, co-pointing, thermal control, data handling, wavelength registration, and suppression of correlated errors.

A quantitative end-to-end mission simulator is essential for optimizing the architecture and establishing a traceable error budget. It should propagate target-star properties, transit geometry, stellar activity, zodiacal and stray-light backgrounds, detector statistics and systematics, fiber-coupling efficiency, WIMS throughput and line-spread function, wavelength calibration, pointing jitter, thermal drift, data gaps, and hierarchical spectral co-addition. The simulator should generate realistic time-series detector data that pass through the same calibration and retrieval pipelines planned for the mission. This framework can then optimize the number and diameter of telescope modules, spectral resolution, channel boundaries, orbit, cadence, target allocation, and mission duration while quantifying biosignature-detection completeness and false-positive risk.

### 7.1 Staged development roadmap

Phase 1 - Component and UPCP validation: characterize CMOS detectors, WSL/WIMS devices, SiC mirrors, calibration sources, and end-to-end recovery of injected ppm-level signals.

Phase 2 - Integrated 1-m module: demonstrate end-to-end throughput, fiber-coupling stability, wavelength stability, pointing sensitivity, environmental performance, and on-sky operation with a flight-representative telescope-spectrograph-detector chain.

Phase 3 - Pathfinder array: operate multiple synchronized modules to validate array-level calibration, timing, co-pointing, data handling, correlated-noise rejection, and hierarchical spectral combination.

Phase 4 - Full Life 2.0 observatory: deploy the replicated array and conduct multi-year transit spectroscopy of the highest-priority nearby Earth 2.0 systems identified by ET and complementary surveys.

## 8. CONCLUSION

Life 2.0 connects the discovery of Earth 2.0 planets with the spectroscopic capability needed to investigate their atmospheres. In geometric collecting area, 900 one-meter telescopes are equivalent to a 30-m aperture; compact WIMS units, sub-electron-noise CMOS detectors, and lightweight SiC mirrors provide a plausible route to implementing that area through replicated modules. The decisive challenge, however, is not aperture alone. The mission must preserve approximately 1 ppm differential stability across many transits and hundreds of independently calibrated optical chains while identifying and suppressing correlated errors that do not average down. Existing demonstrations of WSL throughput and resolving power, integrated on-sky spectroscopy, sub-electron detector noise, and lightweight SiC fabrication provide a promising starting point. The next critical steps are end-to-end UPCP validation with injected sub-ppm signals, a quantitative mission simulator tied to a complete error budget, and a synchronized pathfinder array. If these milestones are achieved, Life 2.0 could transform a difficult monolithic-aperture problem into a scalable precision-spectroscopy observatory capable of testing the atmospheres of nearby Earth 2.0s.

## ACKNOWLEDGMENTS

Jian Ge gratefully acknowledges his Ph.D. advisor, Prof. Roger Angel, for inspiring discussions of the Large Fiber Array Spectroscopic Telescope (LFAST) concept, which helped motivate the distributed-aperture architecture of Life 2.0. This work was supported by China's Space Origins Exploration Program.